\documentclass[aps,prb,twocolumn,showpacs,a4paper,floatfix]{revtex4}
\usepackage{times}
\usepackage[latin1]{inputenc}
\usepackage[english]{babel}
\usepackage[T1]{fontenc}
 \usepackage{graphicx}
\usepackage{epstopdf}
\usepackage[small,FIGTOPCAP]{subfigure}
\usepackage{varioref}
\usepackage{rotating}
\usepackage{mathptmx}
\usepackage{fancyhdr}
\usepackage{amsmath}
\usepackage{amsfonts}
\usepackage{amssymb}
\usepackage{amsbsy}
\usepackage{textcomp}
\usepackage{array}
\usepackage{fixmath}
\usepackage{bm}
\usepackage[active]{srcltx}
\usepackage[colorlinks]{hyperref}
\usepackage[all]{hypcap}

\hypersetup{ pdfcreator=Emiliano Cadelano}

\begin{document}

 \title{Reply to ``Comment on Gap opening in graphene by shear strain''}
 \author{Giulio Cocco,$^1$ Emiliano Cadelano,$^{1}$ Luciano Colombo$^{1}$}

\email[E-mail me at:]{luciano.colombo@dsf.unica.it}
\affiliation{$^1$Department of Physics, University of Cagliari\\
Cittadella Universitaria,
I-09042 Monserrato (Cagliari), Italy}
\date{\today}

\begin{abstract}
In reply to the Comment by Ramasubramaniam regarding our article [Phys. Rev. B {\bf  81}, 241412(R) (2010)] we clarify that our results are indeed valid provided that out-of-plane atomic relaxations are inhibited, as it may occur in the technologically relevant case of supported graphene sheets. We argue that the Comment, while overall interesting, is addressed to the different case of a free standing graphene  monolayer and, therefore, does not  invalidate our conclusions. We also remark that the rippled configuration discussed in the Comment is most likely affected by major size effects and that, contrary to what suggested, our conclusions are not in contrast with the recent work by Pereira {\it et al.} [Phys. Rev. B {\bf 80}, 045401 (2009)].
\end{abstract}
\maketitle

Following the seminal paper by Pereira {\it et al.} \cite{pereira} we recently exploited the concept of strain-induced band structure engineering in graphene through the calculation of its electronic properties under in-plane strain (obtained by uniaxial, shear, and combined uniaxial-shear 
deformations) \cite{cocco}. We showed that by combining shear deformations to uniaxial strains it is possible to affect the gapless electronic structure of graphene by opening a gap as large as $0.9$ eV. The use of a shear component allows for gap opening at a moderate absolute deformation, safely smaller than the graphene failure strain. This result was obtained in absence of out-of-plane deformations (as due, e.g., to bending or rippling), a situation corresponding to configurations where the graphene sheet is supported, i.e. deposited on a suitable substrate \cite{ferrari,ni1,ni2,kim}
 
In the Comment by Ramasubramaniam regarding our article \cite{comment}, it is shown by density-functional tight-binding simulations that the gapless spectrum of a free standing graphene monolayer persists even under shear strain, provided that out-of-plane atomic relaxations (inducing ripples) are allowed. This is consistent with a well known result of continuum mechanics: shear, as well as uniaxial, deformations come with reversible corrugations, whether applied to a free standing elastic membrane \cite{cerda,pellegrino}. Ramasubramanian therefore concluded that  gap opening in graphene under shear deformation (or combined uniaxial and shear deformations) is unlikely to occur, since ripples should cancel the strain effects we found for a flat graphene sheet. 

While overall interesting, we believe that the Comment does not invalidate our conclusions simply because it is addressed to a physical configuration (i.e. a fully flexible monolayer) not equivalent to the one we considered, namely:  a graphene sheet where out-of-plane relaxations are inhibited. The first  and second configurations correspond, respectively, to a suspended and to a supported sample.

In addition, we observe that the ripple geometry of a suspended sheet can be effectively altered via thermal manipulation, up to a complete suppression when temperature is raised to 450-600K, as experimentally found by Bao {\it al.} \cite{bao}. We believe that this result provides another example of
ripple-free, but strained graphene membrane, making our investigation relevant and physically sound even for some suspended samples.

Interesting enough, in Ref.[\onlinecite{bao}] it has been also reported that the measured wavelength of the ripples ranges from 370 to 5000 nm, a much larger value than obtained by Ramasubramanian in his calculation (just few nm). Therefore, real rippled samples are definitely much more flat than described in the Comment. Under this respect, once again we believe that our calculations are meaningful: the larger is the wavelength of the ripples, the better is the approximation of a ripple-free graphene sheet (which locally is basically flat). The large disagreement between the experimental data and the Ramasubramanian calculation about the ripple wavelength is clearly due to the very small size of the periodically-repeated computational sample (a square with length as small as 5 nm). This system, therefore, unlikely represents the case of a real suspended sample and, therefore, the conclusions about the electronic spectrum are questionable.

Finally, we observe that the quotation of the work by Pereira \textit{et al.} \cite{pereira} made in the Comment by Ramasubramanian is not consistent. As a matter of fact, the electronic structure calculations described in Ref.[\onlinecite{pereira}] are performed under the very same assumptions as in our work\cite{cocco}. The fact that the gapless spectrum is found to be robust depends on the actual protocol of deformation (not including out-of-plane relaxations, similarly to our work) considered in that work (where only uniaxial strains are considered), rather than on the formation of ripples (as claimed by Ramasubramanian). Moreover, in a different paper \cite{pereiraPRL} the same Authors further explore the influence of local strain on the electronic structure of graphene. They suggest that the graphene electronics can be controlled  by suitable engineering of local strain profiles,  a perspective which is indeed in nice agreement with our conclusions. 

We acknowledge many useful discussions with Stefano Giordano (Deptartment Physics, University of Cagliari, Italy) and his critical reading of this work.

\bibliographystyle{apsrev}

\end{document}